# Quantifying inhomogeneous magnetic fields at the micron scale using Graphene Hall-Effect sensors

Lionel Petit, Thomas Blon, Benjamin Lassagne

Université de Toulouse, INSA-CNRS-UPS, LPCNO, 135 Av. Rangueil, 31077 Toulouse, France

Corresponding author: lassagne@insa-toulouse.fr

**Abstract:** The response of a graphene Hall-Effect sensor to the inhomogeneous magnetic field generated by a dipole located above it, is investigated numerically at room temperature as a function of the dipole position, its orientation and the conduction regime of the sensor, *i.e.,* diffusive or ballistic. By means of dedicated models, we highlight that the correction factor $\alpha$ frequently used to relate the Hall voltage to the magnetic field averaged over the sensor area can be greatly improved in the high proximity situation enabled by the use of graphene, particularly in the ballistic regime. In addition, it is demonstrated that by fine-tuning the dipole position in the sensor plane, the Hall response becomes highly selective with respect to the dipole orientation. These analyses show that diffusive graphene Hall sensors may be preferred for particle detection, while ballistic ones used as close as possible to a nanomagnet would be preferred for magnetometry. With the support of micromagnetic simulations, the principle of the magnetic hysteresis loop measurement of an isolated nanomagnet is quantitively demonstrated with large signal-to-noise ratio to probe incoherent magnetization reversal. Devices based on specially designed graphene Hall crosses operating in the ballistic regime promise to outperform state-of-the-art ballistic Hall sensors based on semiconductor quantum wells especially for nano-magnetometry.

Quantifying magnetic fields at the micron scale using Hall sensors have been employed extensively since the late 1990s. A wide range of domains were covered, from the imaging of vortices in superconductors [1–3], to the study of ferromagnetic nanoparticles [2,4,5] and biosensing [6–8]. The success of Hall sensors lies mainly on their noninvasiveness, the ease of shrinking them to sub-micrometer scale, the quantitative measurement of the magnetic field and their high sensitivities over a wide range of temperatures and magnetic fields [9,10]. Additionally, the amplitude of the normal component of the magnetic field $B_z$, when homogeneous over the whole sensor area, is directly proportional to the induced Hall voltage $U_H$, i.e., $U_H = B_z I/en$, with $e$ the electronic charge, $I$ the bias current and $n$ the electron density, making it extremely easy to read the sensor output. Although this relationship is no longer valid for an inhomogeneous $B_z$, numerous studies [11–14] demonstrated that, considering the average $B_z$ in the central area of the sensor, the Hall resistance $R_H = U_H/I$ can still be expressed as

$$R_H = \alpha S_I \langle B_z \rangle \tag{1}$$

where $S_I$ is the magnetic field sensitivity ($S_I = 1/ne$) and $\alpha$ is a correction factor that depends on the sensor's geometry, $B_z$ profile and conduction regimes. $\alpha$ has been demonstrated $< 1$ for sensors operated in the diffusive regime [12] and surprisingly $> 1$ for ballistic sensor [14], which allows to amplify the Hall sensor response.

Two-dimensional electron gases (2DEGs) based on semiconductor heterostructures are widely used to fabricate ballistic Hall sensors at low temperatures due to low $n$ and very high electronic mobility $\mu$ [1–5]. But, the discovery of graphene led to a major breakthrough [15] with graphene Hall sensors (GHSs) outperforming any other technology with $S_I > 5\ k\Omega.T^{-1}$ and magnetic field resolution around $50\ nT/\sqrt{Hz}$ at room temperature [16–19]. Additionally, with the current record $\mu > 15\ m^2/(V.s)$ at room temperature [20,21], graphene is the best material for fabricating high-temperature ballistic Hall sensors, paving the way for new developments in magnetic field detection. A first demonstration of the use of GHSs as local magnetometers was made in 2019 with the measurement of an isolated flake of the van der Waals ferromagnet CrBr$_3$ directly deposited on the GHS [22].

In this paper, we present a detailed numerical study of the room temperature GHSs performance for probing the inhomogeneous magnetic stray field of an isolated nanomagnet. Ballistic and diffusive regimes are considered. We show that the ability of GHSs to bring the nanomagnet and sensor closer together below 50 nm, a technological limit for 2DEGs, combined with their low $n$ and high $\mu$, drastically increases $\alpha$. Additionally, the linearity, the dipole sensitivity and selectivity as well as the influence of an applied homogeneous normal magnetic field are explored to demonstrate the ideal conditions for nano-magnetometry.

Hall crosses with a 4-fold symmetry were considered (Fig. 1(a)). The lengths of the four leads were chosen equal to their width $W$ and connected by rounded corners with a radius of curvature $r_c$. The current was injected between the drain set to an electrostatic potential $V_D$ and the grounded source. The two others electrodes, $V_{H_+}$ and $V_{H_-}$, served to measure $U_H$. A gate electrode set at the electrostatic potential $V_g$ and separated from the graphene by an insulating material with an effective dielectric constant of 3.9 and effective thickness of $285\ nm$ was used to tune the graphene Fermi level $E_F$ (see the supplementary material section I). GHSs can be fabricated either using monolayer graphene grown by chemical vapor deposition or exfoliated from a crystal and deposited directly on a SiO$_2$/Si substrate [18,19], or using exfoliated monolayer graphene encapsulated in hexagonal boron nitride (hBN) [16–18]. In the former case, diffusive regimes were observed, while ballistic regimes were observed at room temperature in the latter case [20,21]. To model diffusive GHSs, we used an advanced model based on the Boltzmann formalism that we recently proposed [18] (see the supplementary material section II). We considered graphene with $\mu$ dominated by long range defects [23] with a mean value of $2\ m^2/(V.s)$. To model ballistic GHSs, the so-called semiclassical ballistic billiard model [24] was used. It is particularly adapted here due to the low coherence length expected at room temperature. Thus, the Landauer-Büttiker formula was taken into its finite temperature version [25] :

$$I_j = \frac{ge}{h}\sum_{i\neq j}\left[(\mu_j - \mu_i)\int_{-\infty}^{+\infty}\left(-\frac{\partial f(E)}{\partial E}\right)T_{j\leftarrow i}(E)dE\right] \quad (2)$$

This equation relates the current $I_j$ entering into the device through the electrode $j$ set at an electrochemical potential $\mu_j$ to the electrochemical potential of other electrodes $\mu_i$. $g$ is the graphene valley and spin degeneracy and $T_{j\leftarrow i}(E)$ is the generalized transmission coefficient from the lead $i$ to the lead $j$. The $T_{j\leftarrow i}(E)$ coefficients were computed using the procedure described in the original work of Beenakker and Van Houten [24] (see the supplementary material section III) and Eq. (2) was evaluated numerically (see the supplementary material section III).

As a model for the inhomogeneous $B_z$, we considered the magnetic stray field associated to a dipole having a magnetic moment $m_0 = 5.00 \times 10^6\ \mu_B$, with $\mu_B$ the Bohr magneton, equivalent to the one of a typical $30\ nm$ iron nanocube at saturation [26]. In terms of design, a tradeoff has been found between sufficiently small Hall crosses to detect low $B_z$ inhomogeneities but not too tiny to avoid the significant mobility drop reported in nanoribbons [27] : $W$ and $r_c$ were fixed to 200 and 50 $nm$ respectively. Thus, $W > r_c$ prevents from nonlinear behaviors in the ballistic regime [24]. Simulations were computed considering a graphene Fermi velocity $v_F = 1 \times 10^6 \ m.s^{-1}$ and a dominant thermal doping of $1.6 \times 10^{15} \ m^{-2}$ at the charge neutrality point [18].

Figure 1(b) and 1(c) shows $R_H$ and the longitudinal resistance $R_L$ in both regimes at room temperature as a function of $V_g$ for a homogeneous magnetic field $B_{ho}$ equals to $0.1 \ T$. This is close to the maximum field generated by the dipole of amplitude $m_0$ at a distance of 25 $nm$ corresponding to the smallest distance between the 30 $nm$ nanocube center and the graphene ideally covered by a standard 10 $nm$ thick hBN layer. The maximum $R_L$ is 1.8 times higher in the diffusive regime because of a finite $\mu$, while the two extrema $R_H$ values obtained at the limits of the ambipolar zone differ by less than 15 % between the two regimes. They become equal at zero temperature where only one kind of carrier is present. As a consequence, both regimes can be considered equally sensitive for sensing a homogeneous $B_z$. For the simulations considering inhomogeneous magnetic fields, we choose a doping maximizing $|R_H|$ in the homogeneous case, a condition fulfilled in both regimes for $V_g = 2.3 \ V$ ($n = 2 \times 10^{15} \ m^{-2}$).

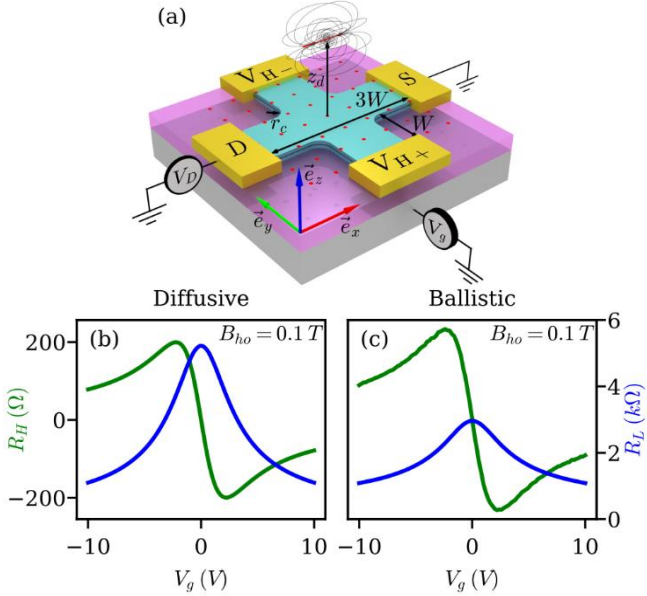

**FIG. 1.** (a) Schematic of a GHS. $R_H$ and $R_L$ in the presence of a homogeneous magnetic field at room temperature as a function of $V_g$ in the (b) diffusive and (c) ballistic regime, respectively.

First, the influence of the dipole position and its magnetic moment orientation on $R_H$ were explored. The magnetic moment was set along either $x$ or $z$ axis, *i.e.,* $m_x$ and $m_z$, while its position $(x_d, y_d, z_d)$ was tuned above the GHS. The $\langle B_z \rangle$ (averaged over the sensor active area defined by the green dotted lines, Fig. 2 line (a)) and $R_H$ (Fig. 2 lines (b)-(c)) values were calculated and displayed as colored pixels corresponding to the different $(x_d, y_d, z_d)$ (the results for *y* orientation are not shown but they can be deduced from $m_x$ results by a clockwise $\pi/2$ rotation followed by a multiplication by -1, see the supplementary material section IV). For dipole-sensor distances as large as $z_d = 100$ nm, (Fig. 2 (a-iii), (b-iii), (c-iii), (a-vi), (b-vi) and (c-vi)), $R_H$ and $\langle B_z \rangle$ maps show similar symmetries and dipole position dependence whatever the considered regime, as approximated by Eq. (1) and as previously observed [11]. However, reducing $z_d$ to 25 nm leads to strong contrasts between $\langle B_z \rangle$ (Fig. 2 (a-i) and (a-iv)) and $R_H$ in the diffusive regime (Fig. 2(b-i) and (b-iv)), and more drastically in the ballistic one (Fig. 2(c-i) and (c-iv)). Very sensitive areas with $R_H$ positive or negative are observed in the $m_x$ configuration while $\langle B_z \rangle \approx 0$ (Fig. 2(c-i)), and unsensitive areas ($R_H = 0$) are observed in the $m_z$ configuration while $\langle B_z \rangle \neq 0$ (Fig. 2(c-iv)). In the diffusive regime, this behavior is only observed near the top left and bottom right corners of the active area for $m_z$ (Fig. 2(b-iv)). This strong dependence of $R_H$ on the dipole position disappears in both regimes gradually with increasing $z_d$.

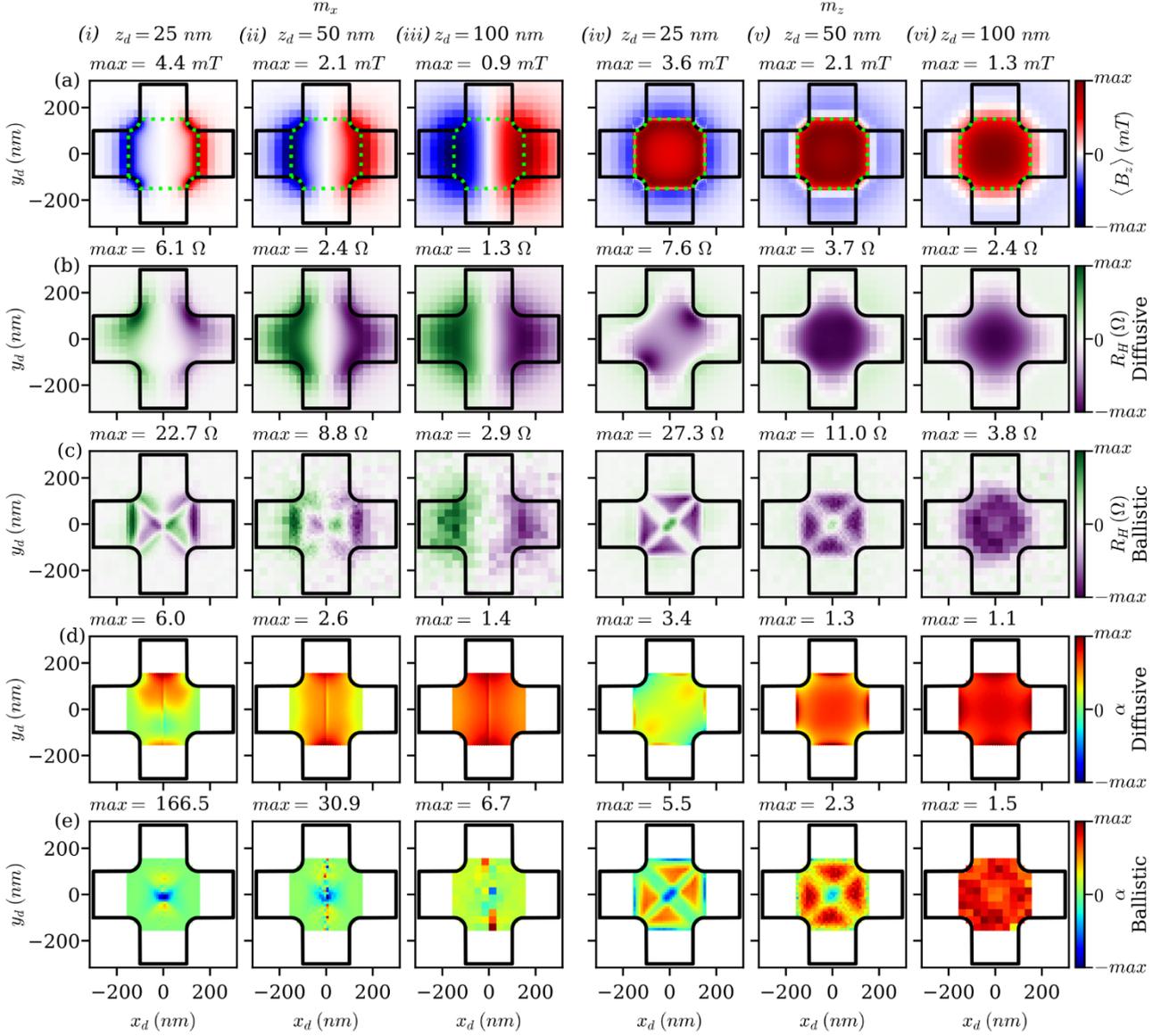

**FIG. 2.** Line (a) color maps of $\langle B_z \rangle$ as a function of $(x_d, y_d)$. Lines (b) and (c), color maps of $R_H$ as a function of $(x_d, y_d)$ for both $m_x$ and $m_z$ dipole orientations and diffusive and ballistic regimes. The dipole was displaced with a step of $10\ nm$ inside the active area and $30\ nm$ outside. Lines (d) and (e), color maps of $\alpha$ as a function of $(x_d, y_d)$ for both $m_x$ and $m_z$ dipole orientations and diffusive and ballistic regimes.

In terms of $R_H$ amplitude, at $25\ nm$ for both orientations, the ballistic regime is almost 4 times more sensitive than the diffusive one despite a nearly equivalent response under homogeneous $B_z$ (Fig. 1(b), (c)). As $z_d$ increases, the $R_H$ amplitude decreases, with a maximal output ratio between ballistic and diffusive signal that drops to 2.4 at $100\ nm$ for $m_x$ and 1.6 for $m_z$. This observation is coherent with the decrease of the $B_z$ inhomogeneity where the ratio is 1.15 for a normal magnetic field (Fig. 1(b, c)). Therefore, to take advantage of the high ballistic sensor's response, it is essential to minimize $z_d$. The $300\ K$ thermal voltage noise limit $\langle \delta V_{noise} \rangle$, evaluated from $\langle \delta V_{noise} \rangle = \sqrt{4 k_B T R_L \Delta f}$ with $\Delta f$ the bandwidth [28], is expected to be in the range of $61\ nV$ and $77\ nV$ for the ballistic and diffusive sensors respectively considering $\Delta f = 100\ Hz$. For a bias current of $1\ \mu A$, Hall voltages of $\pm 20\ \mu V$ and $\pm 5\ \mu V$ for the ballistic and diffusive regime respectively could be expected, which is three orders of magnitude larger than $\langle \delta V_{noise} \rangle$.

The above analysis shows that $R_H$ maps cannot be compared to $\langle B_z \rangle$ ones, especially at small dipole-sensor distances, meaning that Eq. 1 becomes inapplicable as illustrated by Fig. 2 (lines (d) and (e)), where we observe an almost constant, position independent $\alpha$ at large distances, but a strongly position dependent $\alpha$ for small $z_d$, especially in the ballistic regime (Fig. 2 line (e)). In addition, Figure 2 (d-iv) and (e-iv) illustrate that $\alpha_{max}$, the maximum values of $\alpha$, can reach values in specific areas as high as 3.4 (Fig. 2(d-iv)) and 5.5 (Fig. 2(e-iv)) for $m_z$ and $z_d = 25\ nm$ in diffusive and ballistic regimes respectively, which is intrinsically impossible to achieve using 2DEGs Hall sensors. It is worth noting that the very large values of $\alpha_{max}$ obtained for $m_x$ configuration in the ballistic regime are meaningless and correspond to ratio of small $R_H$ amplitudes over near zero $\langle B_z \rangle$. The influences of cross geometry and double cross geometry, as well as others wiring configurations were also investigated and are presented in the supplementary material section IV and V.

In the following, we focus on the performance of GHSs in close proximity with the magnetic dipole, i.e., $z_d = 25\ nm$. The sensor linearity was investigated spatially for a magnetic moment varying continuously over the interval $[-m_0, +m_0]$. For each dipole location, $R_H$ versus moment was fitted by a simple linear regression (intercept was set to $0\ \Omega$). The deduced dipole sensitivity (in $\Omega.\mu_B^{-1}$) is mapped on Figure 3 as well as the sensor linearity given by the determination coefficient $C_D$. Is also displayed the sensor selectivity with respect to the moment orientation by combining the dipole sensitivity along the three axes using the light color addition rules.

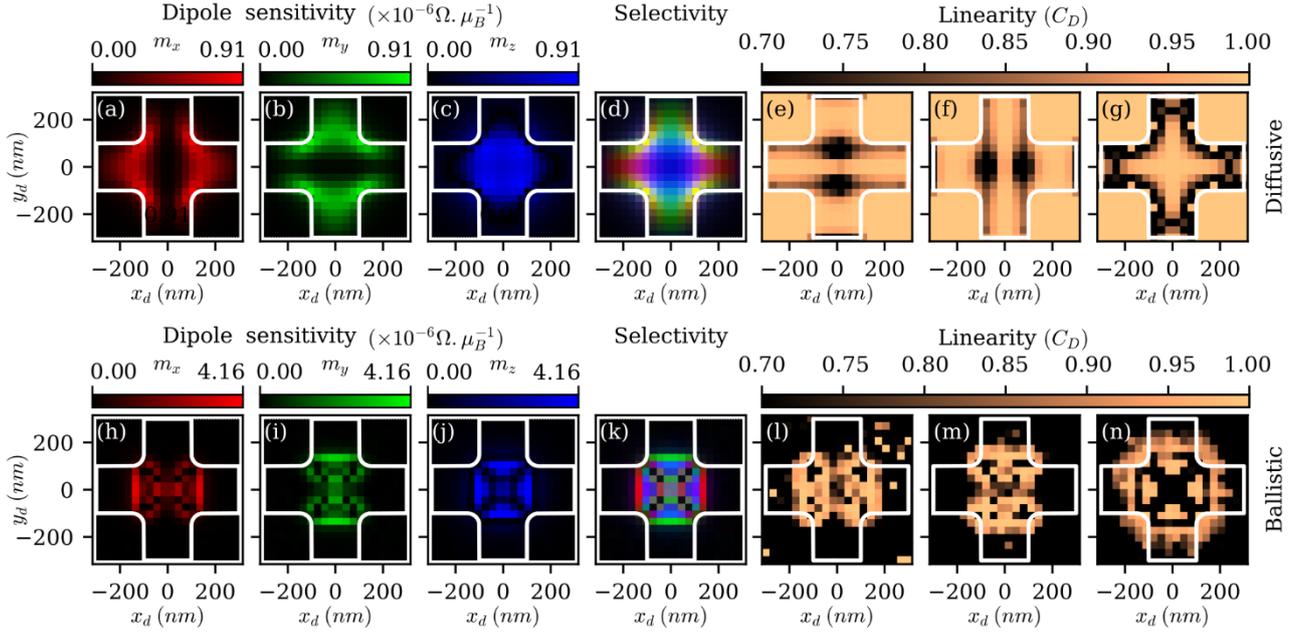

**FIG. 3.** Maps of dipole sensitivities (a-c,h-j), selectivity (d,k) and linearity (e-g,l-n) for diffusive (a-g) and ballistic GHSs (h-n) ($z_d$=25 nm). The selectivity is displayed following the light color addition rules, whereas the three main dipole orientations are plotted as the primary colors.

The sensitivity maps are logically close to the $R_H$ ones (Fig. 2 lines (b)-(c)) in terms of dipole position dependence and amplitude, but some discrepancies appear in areas where linearity is not respected ($C_D < 1$), as shown on the linearity maps (Fig. 3(e)-(g), (l)-(n)). For example, for $m_z$, the corners of the active area display highly non-linear behaviors because they exhibit non symmetrical responses depending on the sign of $m$ and therefore on the $B_z$ one: for $n$ doped sensors, the lower left and upper right corners are more sensitive to $m_z > 0$ and vice versa for $m_z < 0$ (see the supplementary material section VI). It highlights the fact that under inhomogeneous magnetic field, $R_H(x_d, y_d, \vec{m}) \neq -R_H(x_d, y_d, -\vec{m})$. This non-linear response of GHSs for specific dipole positions is characteristic of inhomogeneous magnetic fields and contrasts to the perfect linearity of GHSs observed for homogeneous magnetic fields. The selectivity plot (Fig. 3(d), (k)) shows the ability of GHSs to measure only one specific dipole orientation, that is $m_x$, $m_y$ or $m_z$, displayed as red, green and blue color, respectively. An area equally sensitive to the different orientations is not selective and is shown here in grey shades, like the center of the ballistic GHS in Fig. 3(k), which is also highly non-linear (Fig. 3(n)). In contrast, for a diffusive GHS, the center zone is very sensitive to $m_z$ (Fig. 3(d)) with a perfect linearity (Fig. 3(g)).

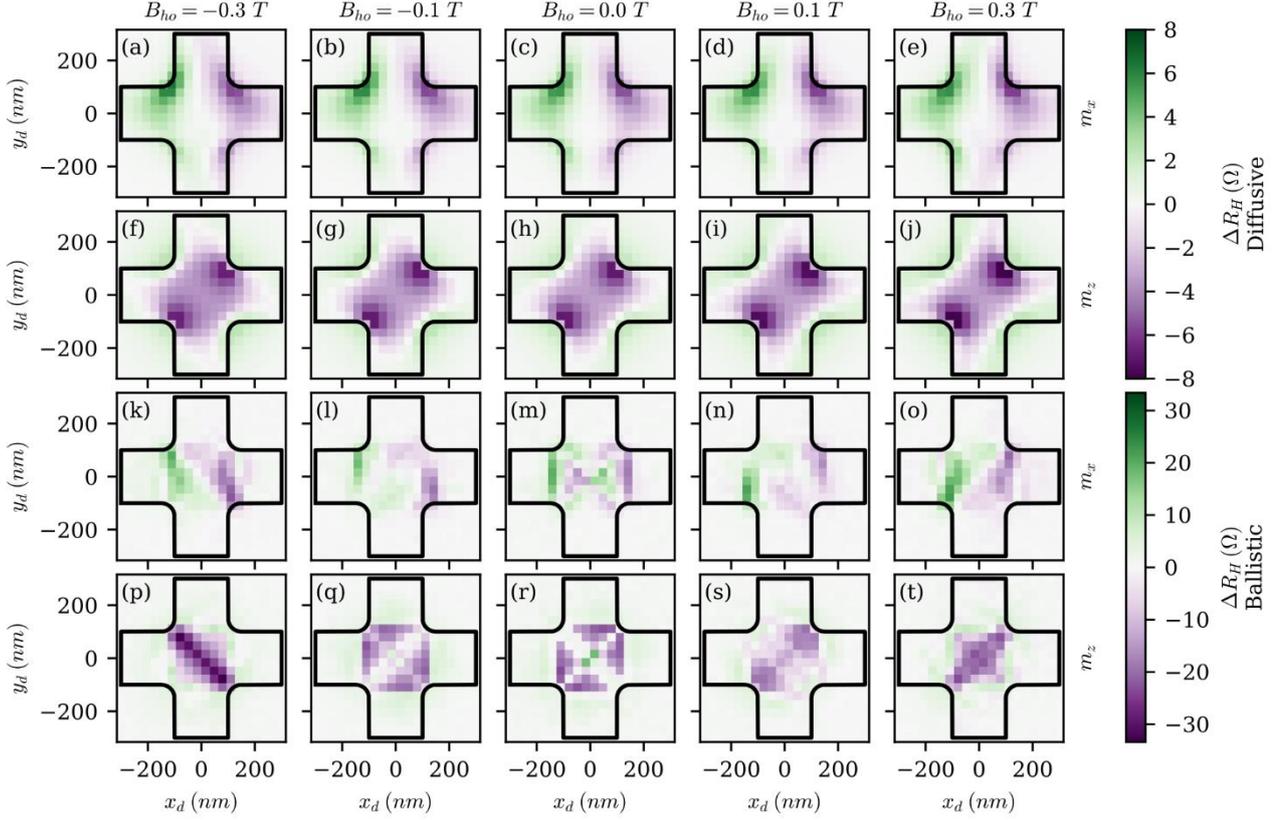

**FIG. 4.** (a-j) $\Delta R_H$ maps in diffusive regime and (k-t) ballistic regime, for different normal $B_{ho}$ ($z_d$=25 nm).

The figure of merit of a ferromagnet is its magnetic hysteresis loop which is obtained by measuring the dependence of its magnetic moment amplitude on a homogeneous magnetic field $B_{ho}$ [1–5,22]. When $m_z$ configuration is investigated, a normal $B_{ho}$ is required, that can have an effect on the dipole sensor response. Figure 4 presents the dipole contribution on the sensor response $\Delta R_H(x_d, y_d, B_{ho}) = R_H(x_d, y_d, B_{ho}) - R_H(B_{ho})$ for different normal $B_{ho}$ amplitudes in the range $\pm 0.3\ T$, at $z_d = 25\ nm$ and constant $m_0$. The $B_{ho}$ influence on $\Delta R_H$ must be kept as low as possible for correctly quantifying the ferromagnet hysteresis loop. In the diffusive regime, $\Delta R_H$ maps are almost independent of $B_{ho}$ whatever the dipole orientation except at the top right and bottom left corners of the active area. The quantity $\langle\ |\Delta R_H| - R_H(x_d, y_d, 0)\rangle$, the magnetic field influence on the dipole sensor response, is less than $0.4\ \Omega$ at $\pm 0.3\ T$, i.e., 10 % of $R_{Hmax}(B_{ho} = 0)$, and decreases with $B_{ho}$. On the contrary, $\Delta R_H$ maps calculated in the ballistic regime are strongly dependent on $B_{ho}$ (Fig. 4(k)-(t))). $\langle\ |\Delta R_H| - R_H(x_d, y_d, 0)\rangle$ is $> 15\ \Omega$ at $B_{ho} = \pm 0.3\ T$, i.e. 65 % of $R_{Hmax}(B_{ho} = 0)$. Therefore, the large modifications of $\Delta R_H$ maps can complicate the quantitative interpretation of the hysteresis loop measurements. It could be noted that this effect could be even more drastic depending on the device geometry and magnetic field amplitude. Whereas the present cross allows carriers to flow through the whole device due to their suitable cyclotron radius ($W = 200\ nm$ combined with the considered doping and magnetic field amplitude), higher $B_{ho}$, wider crosses or smaller doping would increase guiding behavior [24] due to a cyclotron radius lower than $W$. As a consequence, the surface area of insensitive zones in the device would increase, especially in the central part of the active area [11]. In such conditions, the $\Delta R_H$ maps alterations would be even more pronounced.

The above analyses show that, depending on the targeted application, diffusive or ballistic GHSs are preferable. For particle detection, as frequently used in biosensing, diffusive GHSs may be preferred as $R_H$ does not depend so much on the dipole position. For magnetometry, ballistic GHSs as close as possible with the nanomagnet are clearly dedicated for high output signal, although this approach requires fine nanomagnet positioning. For illustration purposes, $R_H$ loop hysteresis induced by the magnetic hysteresis of an $30\ nm$ iron nanocube was simulated in combination with micromagnetic simulations (see the supplementary material section VII). In order to maximize $R_H$, the nanocube was positioned at $25\ nm$ above the GHS at an in-plane position sharing high sensitivity, linearity and selectivity (Fig. 4(k)-(l)) and Fig. 5(a)). This position is highly $m_x$ sensitive, as a consequence $B_{ho}$ was applied in the $x$ direction to avoid any $R_H$ alteration.

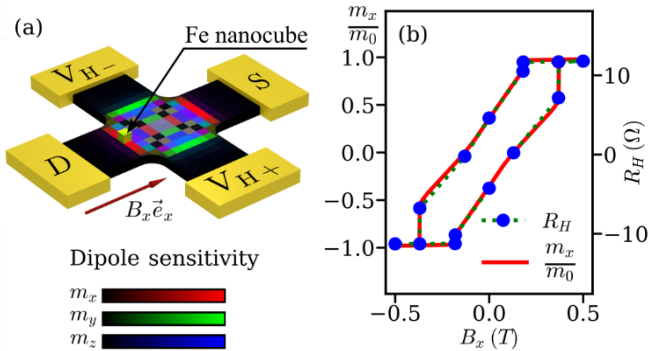

**FIG. 5.** (a) Fe nanocube on top of a hBN encapsulated GHS. (b) Magnetic hysteresis loop of the nanocube (red) and corresponding $R_H$ computed using the ballistic model (green dotted line with blue markers).

Figure 5(b) shows the $R_H$ (green doted curve with blue markers) and magnetization hysteresis loops along the $x$ axis (red curve). The $R_H$ loop captures perfectly the magnetic hysteresis loop. Note that we have intentionally chosen a nanomagnet with a hysteresis loop that differs from the usual square loop, in this case a sheared cycle with a vortex-assisted magnetization reversal in the linear part [26]. This incoherent magnetic reversal induces local hysteresis features as small as few $\Omega$ which would induce voltages at 1 $\mu A$ well above $\langle \delta V_{noise} \rangle$, *i.e.*, a signal-to-noise ratio of a few tens.

In conclusion, we investigated new possibilities offered by graphene for Hall sensors at room temperature. We simulated the Hall response of diffusive and ballistic GHSs to the magnetic stray field of a magnetic dipole having different orientations and positions. It appears that $\alpha$ is strongly enhanced in the situation of close proximity, especially in the ballistic regime, with the counterpart of presenting a Hall response strongly dependent on position and on a normal applied magnetic field. However, by fine-tuning the experimental conditions, GHSs turns out to be highly sensitive fluxmeters up to room temperature, outperforming current ballistic Hall sensors based on 2DEGs, which only operate at low temperatures. This result opens up new avenues for Hall sensors from micro-magnetometry to nano-magnetometry.

The data that support the findings of this study are available from the corresponding author upon reasonable request.

**ACKNOWLEDGMENTS**

This study has been supported through the EUR grant NanoX n° ANR-17-EURE-0009 in the framework of the "Programme des Investissements d'Avenir".

[1] A. K. Geim, I. V. Grigorieva, S. V. Dubonos, J. G. S. Lok, J. C. Maan, A. E. Filippov, and F. M. Peeters, *Phase Transitions in Individual Sub-Micrometre Superconductors*, Nature **390**, 259 (1997).
[2] A. K. Geim, S. V. Dubonos, J. G. S. Lok, I. V. Grigorieva, J. C. Maan, L. T. Hansen, and P. E. Lindelof, *Ballistic Hall Micromagnetometry*, Appl. Phys. Lett. **71**, 2379 (1997).
[3] S. Pedersen, G. R. Kofod, J. C. Hollingbery, C. B. Sørensen, and P. E. Lindelof, *Dilation of the Giant Vortex State in a Mesoscopic Superconducting Loop*, Phys. Rev. B **64**, 104522 (2001).
[4] L. Theil Kuhn, A. K. Geim, J. G. S. Lok, P. Hedegård, K. Ylänen, J. B. Jensen, E. Johnson, and P. E. Lindelof, *Magnetisation of Isolated Single Crystalline Fe-Nanoparticles Measured by a Ballistic Hall Micro-Magnetometer*, Eur. Phys. J. D **10**, 259 (2000).
[5] A. D. Kent, S. V. Molnár, S. Gider, and D. D. Awschalom, *Properties and Measurement of Scanning Tunneling Microscope Fabricated Ferromagnetic Particle Arrays (Invited)*, Journal of Applied Physics **76**, 6656 (1994).
[6] P.-A. Besse, G. Boero, M. Demierre, V. Pott, and R. Popovic, *Detection of a Single Magnetic Microbead Using a Miniaturized Silicon Hall Sensor*, Appl. Phys. Lett. **80**, 4199 (2002).
[7] G. Mihajlović, P. Xiong, S. von Molnár, K. Ohtani, H. Ohno, M. Field, and G. J. Sullivan, *Detection of Single Magnetic Bead for Biological Applications Using an InAs Quantum-Well Micro-Hall Sensor*, Appl. Phys. Lett. **87**, 112502 (2005).

[8] L. Di Michele, C. Shelly, J. Gallop, and O. Kazakova, *Single Particle Detection: Phase Control in Submicron Hall Sensors*, J. Appl. Phys. **108**, 103918 (2010).

[9] R. S. Popović, *Hall Effect Devices*, 2. ed (Institute of Physics Publ, Bristol, 2004).

[10] G. Boero, M. Demierre, P.-. A. Besse, and R. S. Popovic, *Micro-Hall Devices: Performance, Technologies and Applications*, Sensors and Actuators A: Physical **106**, 314 (2003).

[11] F. M. Peeters and X. Q. Li, *Hall Magnetometer in the Ballistic Regime*, Appl. Phys. Lett. **72**, 572 (1998).

[12] I. S. Ibrahim, V. A. Schweigert, and F. M. Peeters, *Diffusive Transport in a Hall Junction with a Microinhomogeneous Magnetic Field*, Phys. Rev. B **57**, 15416 (1998).

[13] S. J. Bending, K. V. Klitzing, and K. Ploog, *Two-Dimensional Electron Gas as a Flux Detector for a Type-II Superconducting Film*, Phys. Rev. B **42**, 9859 (1990).

[14] M. Cerchez and T. Heinzel, *Correction Factor in Nondiffusive Hall Magnetometry*, Appl. Phys. Lett. **98**, 232111 (2011).

[15] K. S. Novoselov, *Electric Field Effect in Atomically Thin Carbon Films*, Science **306**, 666 (2004).

[16] B. T. Schaefer, L. Wang, A. Jarjour, K. Watanabe, T. Taniguchi, P. L. McEuen, and K. C. Nowack, *Magnetic Field Detection Limits for Ultraclean Graphene Hall Sensors*, Nat Commun **11**, 4163 (2020).

[17] J. Dauber, A. A. Sagade, M. Oellers, K. Watanabe, T. Taniguchi, D. Neumaier, and C. Stampfer, *Ultra-Sensitive Hall Sensors Based on Graphene Encapsulated in Hexagonal Boron Nitride*, Appl. Phys. Lett. **106**, 193501 (2015).

[18] L. Petit, T. Fournier, G. Ballon, C. Robert, D. Lagarde, P. Puech, T. Blon, and B. Lassagne, *Performance of Graphene Hall Effect Sensors: Role of Bias Current, Disorder, and Fermi Velocity*, Phys. Rev. Applied **22**, 014071 (2024).

[19] A. Tyagi, L. Martini, Z. M. Gebeyehu, V. Mišeikis, and C. Coletti, *Highly Sensitive Hall Sensors Based on Chemical Vapor Deposition Graphene*, ACS Appl. Nano Mater. acsanm.3c03920 (2023).

[20] L. Wang et al., *One-Dimensional Electrical Contact to a Two-Dimensional Material*, Science **342**, 614 (2013).

[21] D. G. Purdie, N. M. Pugno, T. Taniguchi, K. Watanabe, A. C. Ferrari, and A. Lombardo, *Cleaning Interfaces in Layered Materials Heterostructures*, Nat Commun **9**, 5387 (2018).

[22] M. Kim et al., *Micromagnetometry of Two-Dimensional Ferromagnets*, Nat Electron **2**, 457 (2019).

[23] T. Stauber, N. M. R. Peres, and F. Guinea, *Electronic Transport in Graphene: A Semi-Classical Approach Including Midgap States*, Phys. Rev. B **76**, 205423 (2007).

[24] C. W. J. Beenakker and H. van Houten, *Billiard Model of a Ballistic Multiprobe Conductor*, Phys. Rev. Lett. **63**, 1857 (1989).

[25] S. Datta, *Electronic Transport in Mesoscopic Systems* (Cambridge university press, Cambridge, 1995).

[26] F. J. Bonilla, L.-M. Lacroix, and T. Blon, *Magnetic Ground States in Nanocuboids of Cubic Magnetocrystalline Anisotropy*, Journal of Magnetism and Magnetic Materials **428**, 394 (2017).

[27] Yinxiao Yang and R. Murali, *Impact of Size Effect on Graphene Nanoribbon Transport*, IEEE Electron Device Lett. **31**, 237 (2010).

[28] A. Oral, S. J. Bending, and M. Henini, *Real-time Scanning Hall Probe Microscopy*, Appl. Phys. Lett. **69**, 1324 (1996).

# Supplementary Material:

# Quantifying inhomogeneous magnetic fields at the micron scale using Graphene Hall-Effect sensors

**Lionel Petit, Thomas Blon, Benjamin Lassagne**

Université de Toulouse, INSA-CNRS-UPS, LPCNO, 135 Av. Rangueil, 31077 Toulouse, France

Corresponding author: lassagne@insa-toulouse.fr

## I. Electrostatic doping

The position of the Fermi level $E_F$ was determined by the neutrality equation [1–4]

$$Q_g + Q_{gr}(E_F) = 0 \qquad (s1)$$

Where $Q_{gr}(E_F) = -e\big(n(E_F) - p(E_F)\big)$ is the electric charge per surface unit in graphene and $n(E_F)$ and $p(E_F)$ are the electron and hole density respectively determined using the two following relationships

$$n(E_F) = \int_{E_{CNP}}^{\infty} f(E) \frac{2|E - E_{CNP}|}{\pi(\hbar v_F)^2} dE \qquad (s2a)$$

$$p(E_F) = \int_{-\infty}^{E_{CNP}} \big(1 - f(E)\big) \frac{2|E - E_{CNP}|}{\pi(\hbar v_F)^2} dE \qquad (s2b)$$

With $E$, the electron energy, $E_{CNP}$, the energy at the charge neutrality point (CNP) and $v_f$ the Fermi velocity. $f(E) = 1/\big(1 + exp((E - E_F)/k_B T)\big)$ is the Fermi-Dirac distribution with $T$ the temperature. In this study we considered $v_F = 1 \times 10^6 \, m.s^{-1}$, but it is worth to note that higher values of $v_F$ have been reported, especially for graphene encapsulated with hexagonal Boron Nitride [4–6] allowing therefore lower thermal doping and higher magnetic field sensitivities [4].

$Q_g(x,y)$ is the electric charge per surface unit on the gate electrode and is given by

$$Q_g(x,y) = C_{geo}(V_g - V_{gr}) \quad (s3)$$

with $C_{geo}$ the geometric capacitance per surface unit and $V_{gr}$ the graphene electrostatic potential. In our study, low $I$ was assumed, meaning that $V_{gr}$ is simply equals to the shift of the Fermi level, $V_{gr} = (E_f - E_{CNP})/e$ [4].

## II. Diffusive Model

To model diffusive graphene Hall sensors, we used an advanced model based on the Boltzmann formalism that we recently proposed [4]. In our modelling, we considered graphene with an electronic mobility $\mu$ dominated by long range defects [7] with a mean value fixed to $2\ m^2/(V.s)$. It means that $\mu$ is independent of the charge carrier doping. Additionally, as previously reported [4], the electron-hole pair generation and recombination were modelled with a simple linear relationship such that the recombination-generation rate of carriers $R$ written as

$$R = k_r(np - n_{eq}p_{eq}) \quad (s4)$$

with a recombination parameter $k_r = 10^{-4}\ m^2.s^{-1}$ (Petit) and $n_{eq}$, $p_{eq}$ the equilibrium electron and hole densities [4].

## III. Ballistic Model

To model ballistic graphene Hall sensors, we have used the semiclassical billiard model of Beenakker and van Houten [8]. Thus, the Landauer-Büttiker formula was taken into its finite temperature version [9].

$$I_j = \frac{ge^2}{h} \sum_{i \neq j} \left[ (\mu_j - \mu_i) \int_{-\infty}^{+\infty} \left( -\frac{\partial f(E)}{\partial E} \right) T_{j \leftarrow i}(E) dE \right] \quad (s5)$$

This equation relates the current $I_j$ entering into the device through the electrode $j$ fixed at an electrochemical potential $\mu_j$ to the electrochemical potential of other electrodes $\mu_i$. $g$ is the graphene valley and spin degeneracy and $T_{j \leftarrow i}(E)$ is the generalized transmission coefficient from the lead $i$ to the lead $j$. They are defined such that

$$\sum_j T_{j \leftarrow i}(E) = M_i \quad (s6)$$

with $M_i$ the number of modes present in the contact $i$ [9]. Despite its intrinsic discrete nature, the number of modes $M_i$ is taken as continuous in this semiclassical model to preserve the equality $R_{ij}(B) = R_{ji}(-B)$ where $R_{ij}$ is the resistance tensor element. The $T_{j \leftarrow i}(E)$ coefficients were computed through Monte Carlo simulations by evaluating classical trajectories of charges. With a uniform probability entrance along the contact and an initial angle $\theta_c$ with respect to the contact normal following the distribution $P(\theta_c) = \frac{cos(\theta_c)}{2}$, we computed step by step the positions and speeds of charge carriers. For a certain energy $E$, the evaluation of a large number of charge trajectories from the contact labeled $i$ allows to derive the $T_{j \leftarrow i}(E)$ coefficient following through

$$T_{j \leftarrow i}(E) = N_{j \leftarrow i} \times M_i \Big/ \sum_j N_{j \leftarrow i} \quad (s7)$$

Where $N_{j \leftarrow i}$ is the number of charges absorbed in the $j$ contact coming from the $i$ contact. The evaluation of the Eq. (s5) was done numerically by restricting the whole energy domain to a set of 40 energies in the range $[-6.5k_BT, 6.5k_BT]$ around the Fermi energy. This discretization presented in Fig. S1(c) was chosen as a tradeoff between precision and calculation speed since no significant modification was noticeable by further refinement as illustrated on Fig. S1(f)-(h), where the Hall resistance $R_H$, the longitudinal $R_L$ and bend resistance $R_B$ were calculated for a homogeneous magnetic field and plotted as a function of the gate voltage. In the presence of an inhomogeneous magnetic field, we evaluated the $T_{j\leftarrow i}(E)$ coefficient by computing the trajectories of typically $10^5$ charges per contact and per energy group. This procedure was repeated 10 times to evaluate the statistical error.

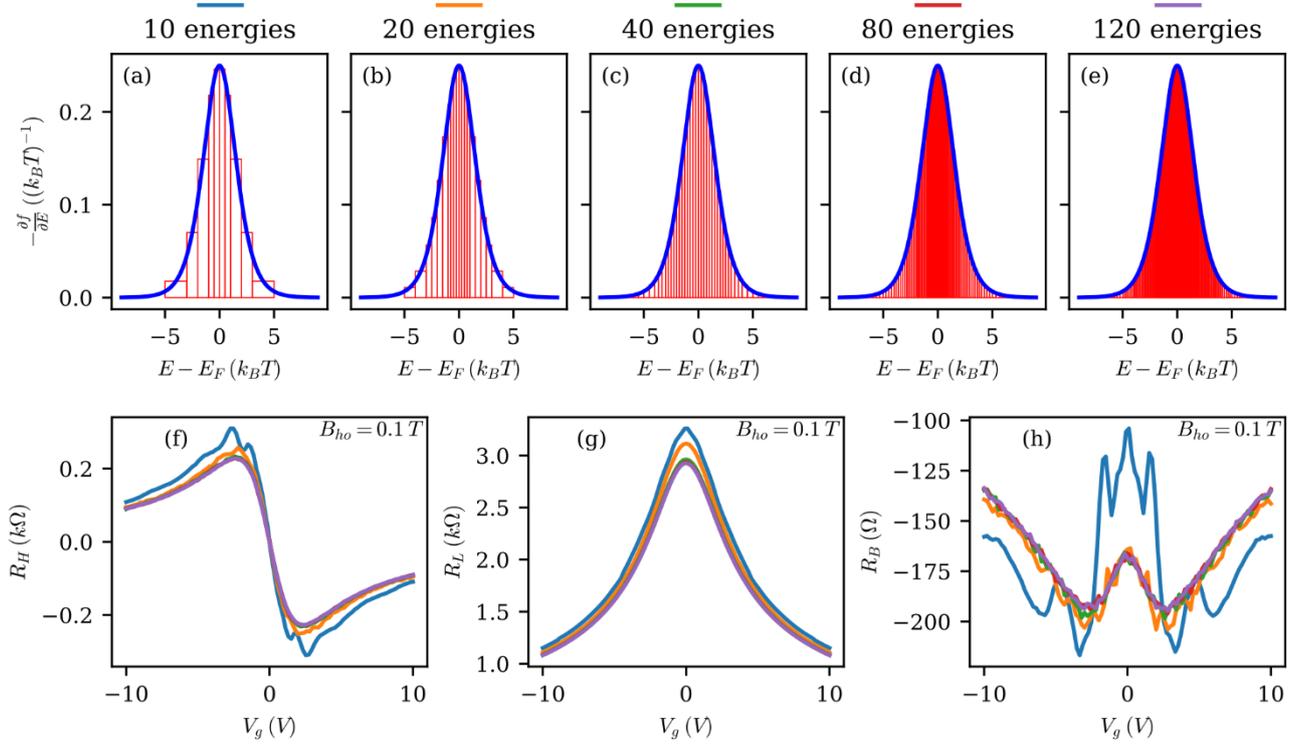

**FIG. S1.** Fermi-Dirac distribution discretization influence on main resistances. (a-e) Different Fermi-Dirac distribution discretization considered. (f-h) Main characteristics as a function of the gate voltage at fixed magnetic field of $0.1\ T$.

# IV. Geometry influence

## A. Curvature radius

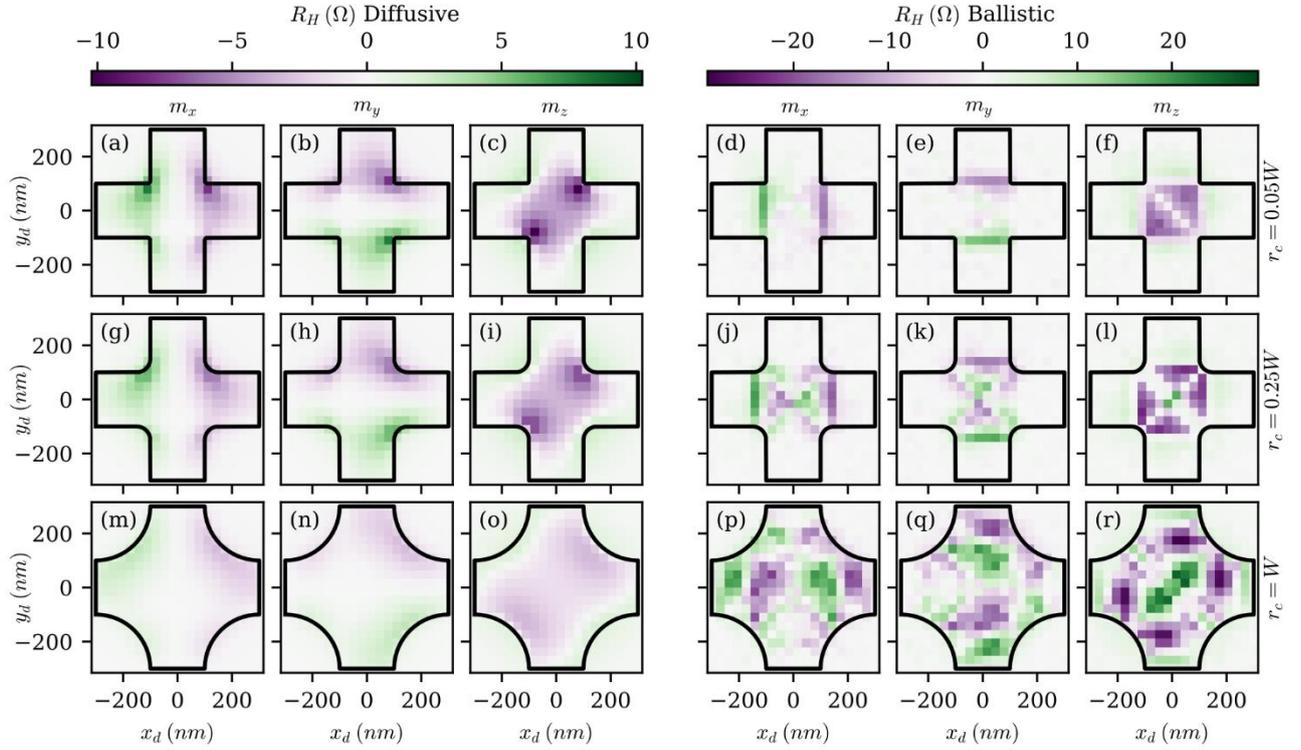

**FIG. S2.** Corner radius influence on the $R_H$ maps in the (a-c, g-i, m-o) diffusive and (d-f, j-l, p-r) ballistic regimes. The magnetic dipole of $5.00 \times 10^6 \, \mu_B$ is placed $25 \, nm$ above the graphene plane. (a-f), (g-l) and (m-r) concern corner radii of $10 \, nm$, $50 \, nm$ and $200 \, nm$ respectively.

## B. Cross size

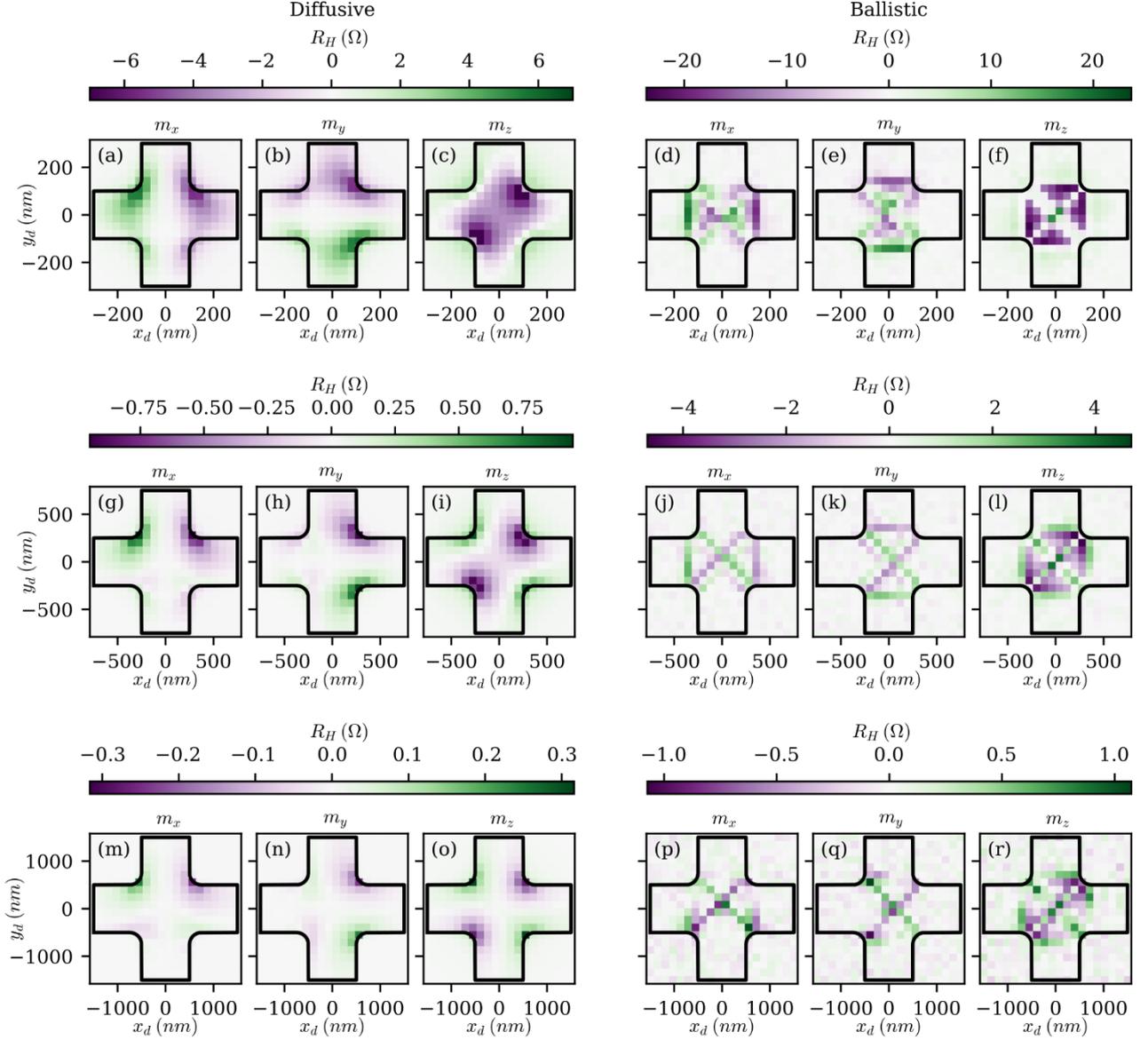

**FIG. S3.** $R_H$ maps for different Hall crosses width, $W = 200\ nm$, $W = 500\ nm$ and $W = 1000\ nm$ for (a-f), (g-l) and (m-r) respectively. The ratio of radius curvature to arm width was kept constant at $r_c/W = 0.25$.

## C. Double cross investigations

The double cross geometry is frequently used to remove by simple subtraction the contribution of an applied homogeneous perpendicular magnetic field [10–13].

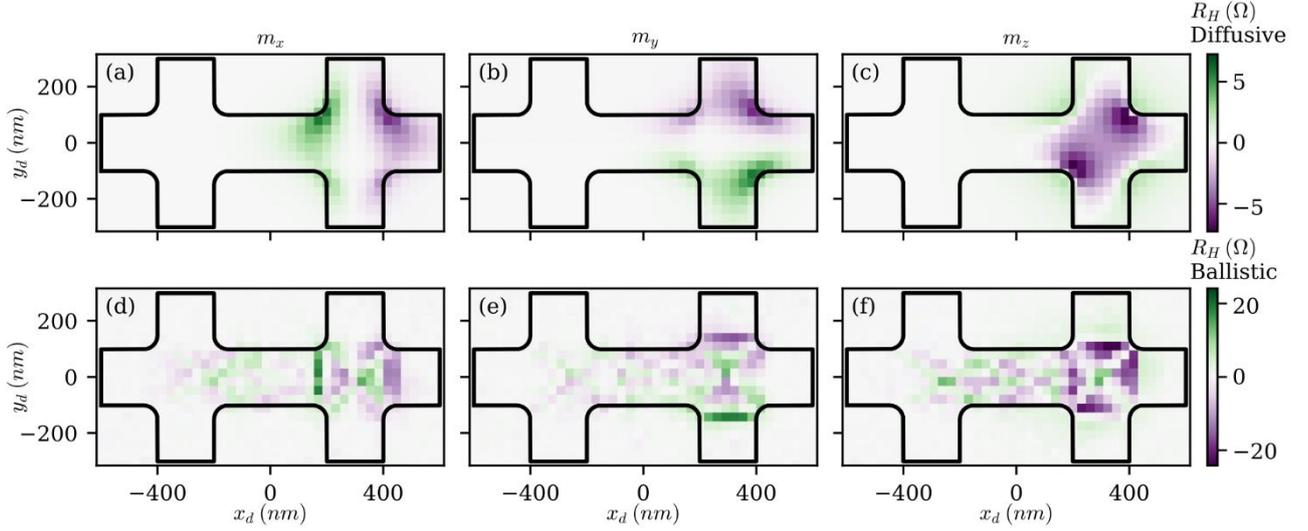

**FIG. S4.** Hall resistance in a double cross evaluated on the right junction as a function of the dipole position. (a, c, e) Diffusive regime for $m_x$, $m_y$ and $m_z$ respectively. (b, d, f) Ballistic regime for $m_x$, $m_y$ and $m_z$ respectively.

Figure S4 shows $R_H$ on the right junction as a function of the dipole position for $m_x$, $m_y$ and $m_z$, and both regimes at $z_d = 25\ nm$. For all cases, the maximum $R_H$ corresponds to the active area of the right cross. For diffusive regime (Fig. S4(a)-(c)), $R_H$ maps are similar to those obtained for the single cross (Fig. S3(a)-(c)). On the contrary, in ballistic regime (Fig. S4(d)-(f)), a modification of the $R_H$ maps with an amplitude decrease on the left side of the right active area is observed (Fig. S3(d)-(f)). This is accompanied by a non-negligible sensitive area that extends across the entire left arm and up to the left cross. Similar simulations were performed with one cross with longer arms (Fig. S5). The sensitive areas were not affected proving that the $R_H$ modification is due to the additional empty ballistic cross which causes a modification of the injection distribution in the right cross (Fig. S6). It involves that for ballistic sensors operating under inhomogeneous magnetic field, the complete geometry has to be considered. One way to simplify the problem is to decouple the right cross from the left one by largely increasing the distance between Hall crosses or by placing a metallic electrode between the two crosses to reset the electron distribution. The drawback is the potentially high contact resistance added.

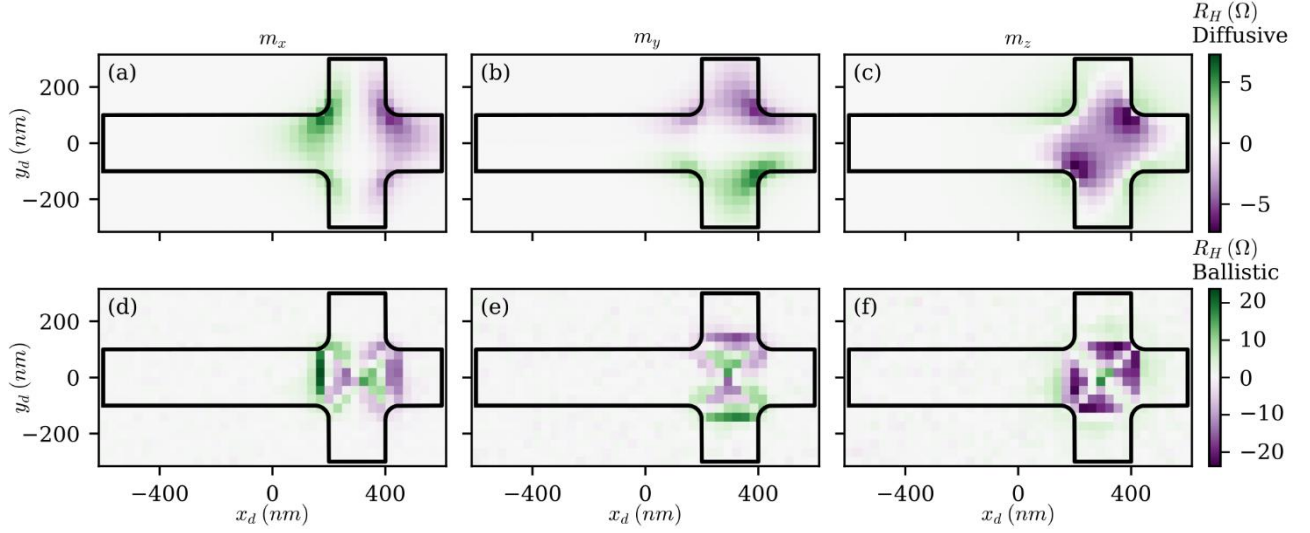

**FIG. S5.** Hall resistance as a function of magnetic dipole position in a non-symmetric Hall cross (a-c) in diffusive regime and (d-f) in ballistic regime.

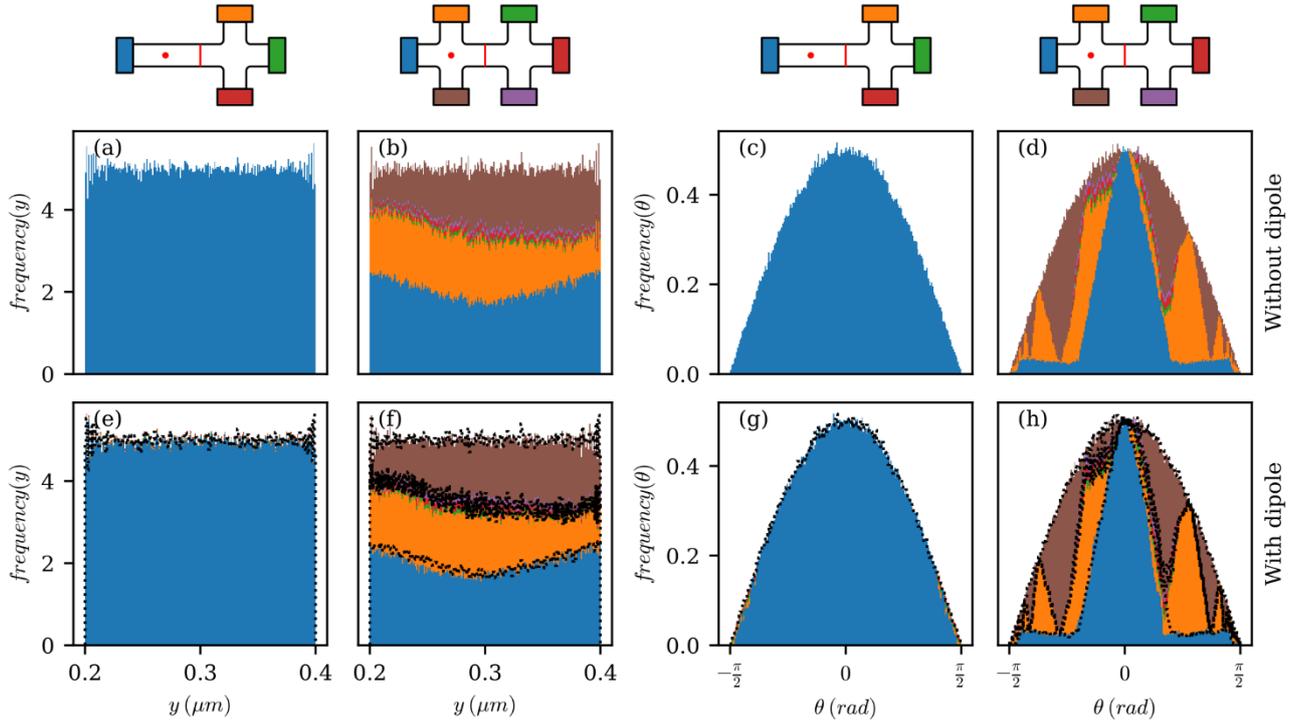

**FIG. S6.** Position (a, b, e, f) and angular (c, d, g, h) distribution at the entrance of the right cross (vertical red line). Figures (a-d) represents these distributions in the absence of a magnetic field while (e-h) correspond to the distribution in the presence of a magnetic dipole of $5.00 \times 10^6 \; \mu_B$ oriented along the $z$ axis and located in the center of the left part of the bridge $25 \; nm$ above the graphene plane (red dot). The black dotted line shows the reference

distribution. (a, e, c, g) present the case of a double cross bridge while (c, d, g, h) present the case of a simple dissymmetric Hall cross.

## V. Different wiring configurations

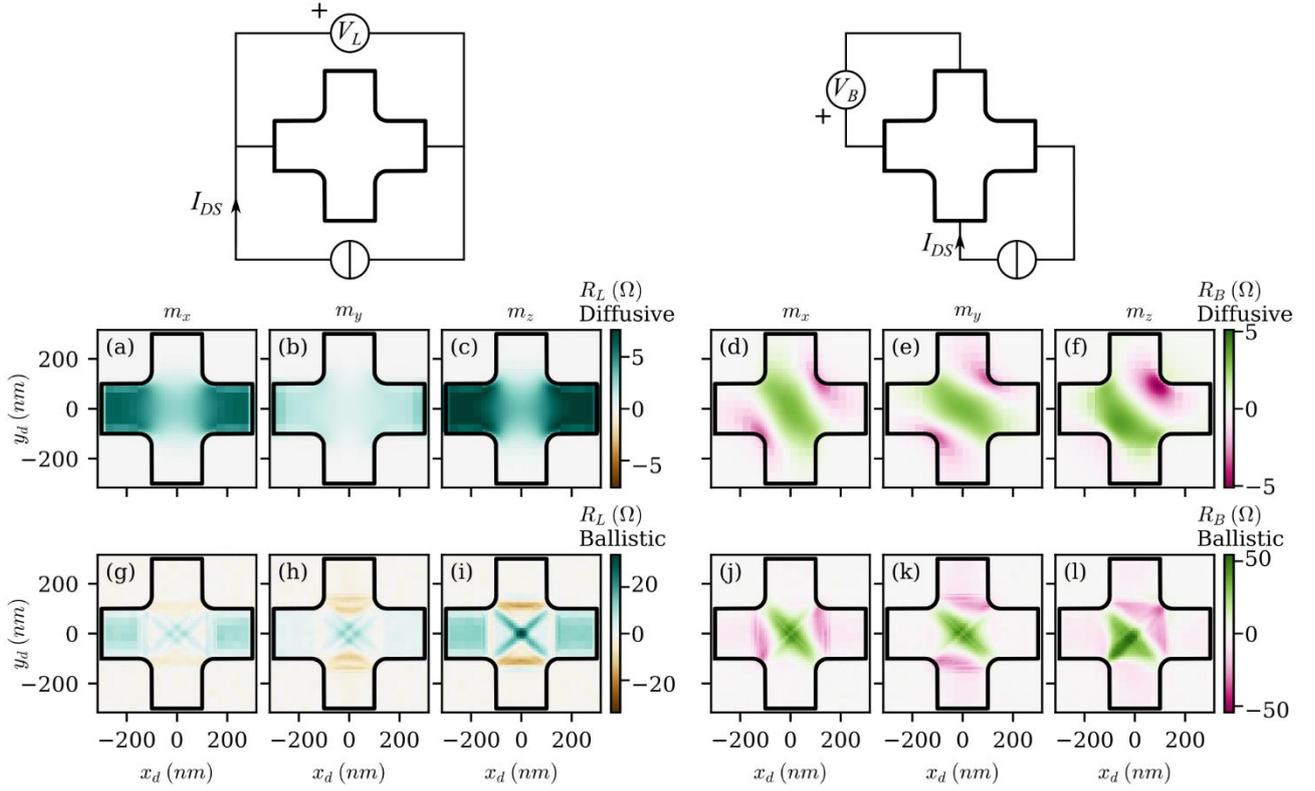

**FIG. S7.** Magnetic dipole position influence on longitudinal resistance (a-c, g-i) and bend resistance (d-f, j-l). The dipole of $5.00 \times 10^6 \, \mu_B$ is located at $25 \, nm$ above the graphene plane, and the gate voltage is set at $2.3 \, V$ on a $285 \, nm$ thick oxide.

# VI. $R_H$ non linearities

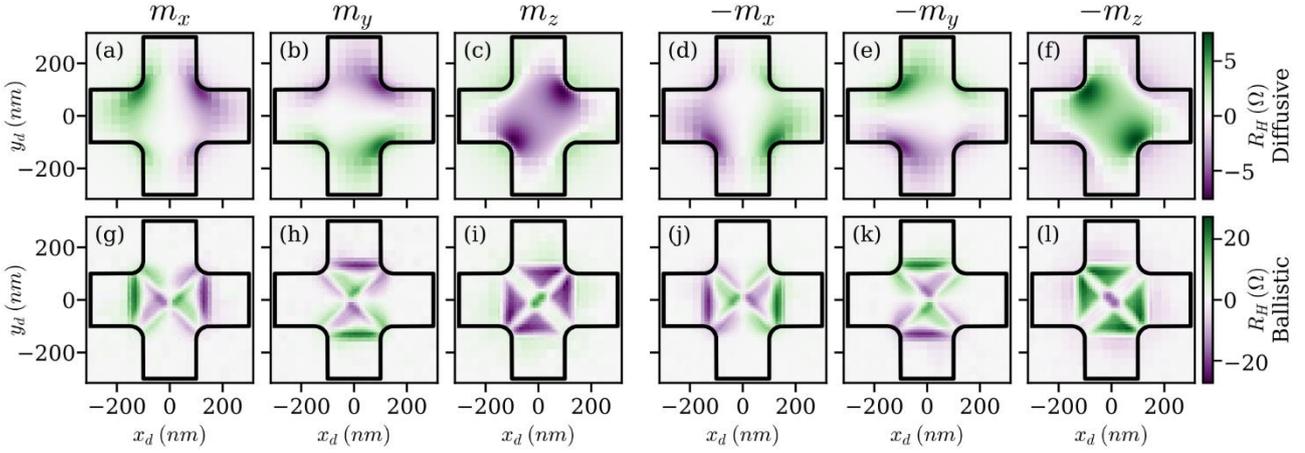

**FIG. S8.** Hall resistance as a function of magnetic dipole position located 25 nm above the surface (a-f) in diffusive regime and (g-l) in ballistic regime. Main moment orientations are considered including negative ones.

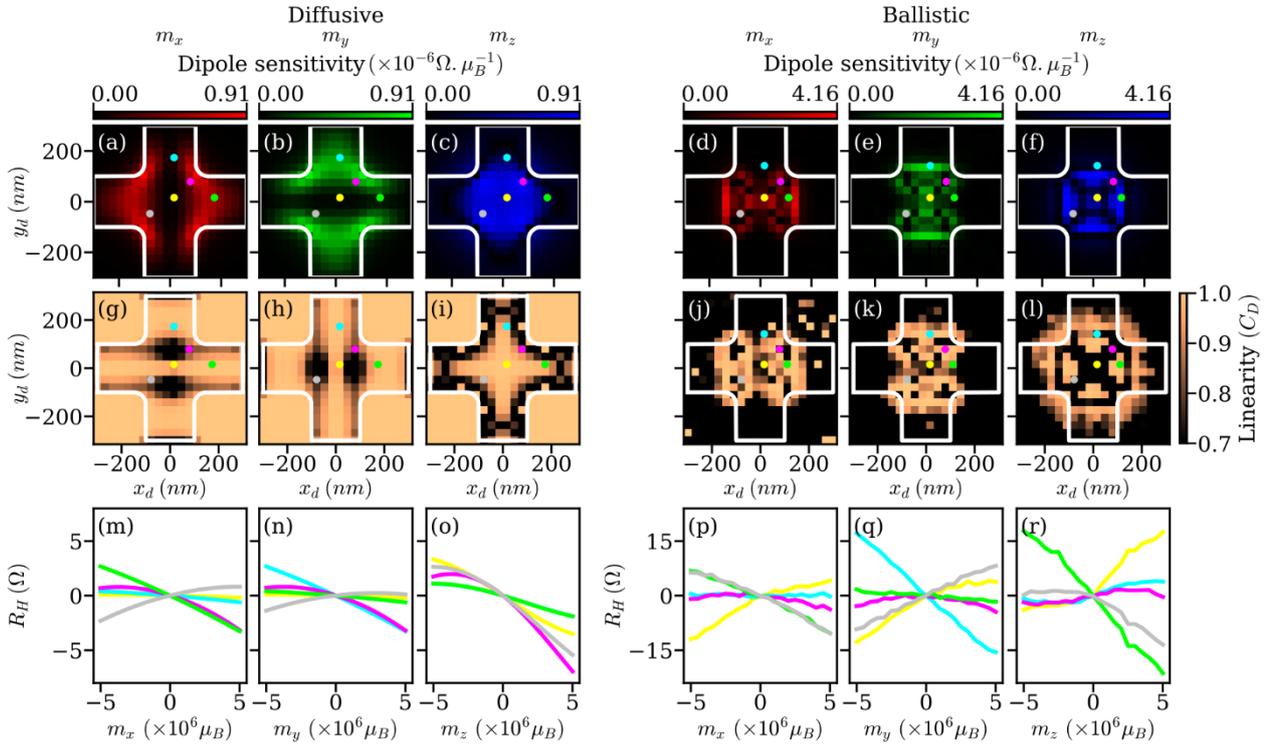

**FIG. S9.** (a-f) Dipole sensitivity and (g-l) linearity of the sensor with respect to dipole amplitude. (m-r) Evolution of $R_H$ with respect to dipole amplitude for different dipole positions.

# VII. Micromagnetic simulation parameters

To investigate the graphene Hall sensor performance to probe the magnetic properties of an individual nano-magnet at room temperature, we computed with OOMMF micromagnetic software the magnetic hysteresis loop of an 30 $nm$ iron nanocube (Fig. S10). The magnetic parameters of the nanocube have been set at the values of bulk iron: saturation magnetization $M_s = 1.72 \times 10^6 \, A.m^{-1}$, cubic magnetocrystalline anisotropy $K_1 = 4.8 \times 10^4 \, J.m^{-3}$ and exchange stiffness to $A = 2.1 \times 10^{-11} \, J.m^{-1}$ [14]. The nanocube was discretized in 3D cuboid cells with 2 nm side length, *i.e.*, smaller than the exchange length in Fe ($l_{ex} = (A/\mu_0 M_s^2)^{1/2} = 2.4$ nm). The micromagnetic simulations were conducted in a total space of $80 \times 80 \times 80 \, nm^3$.

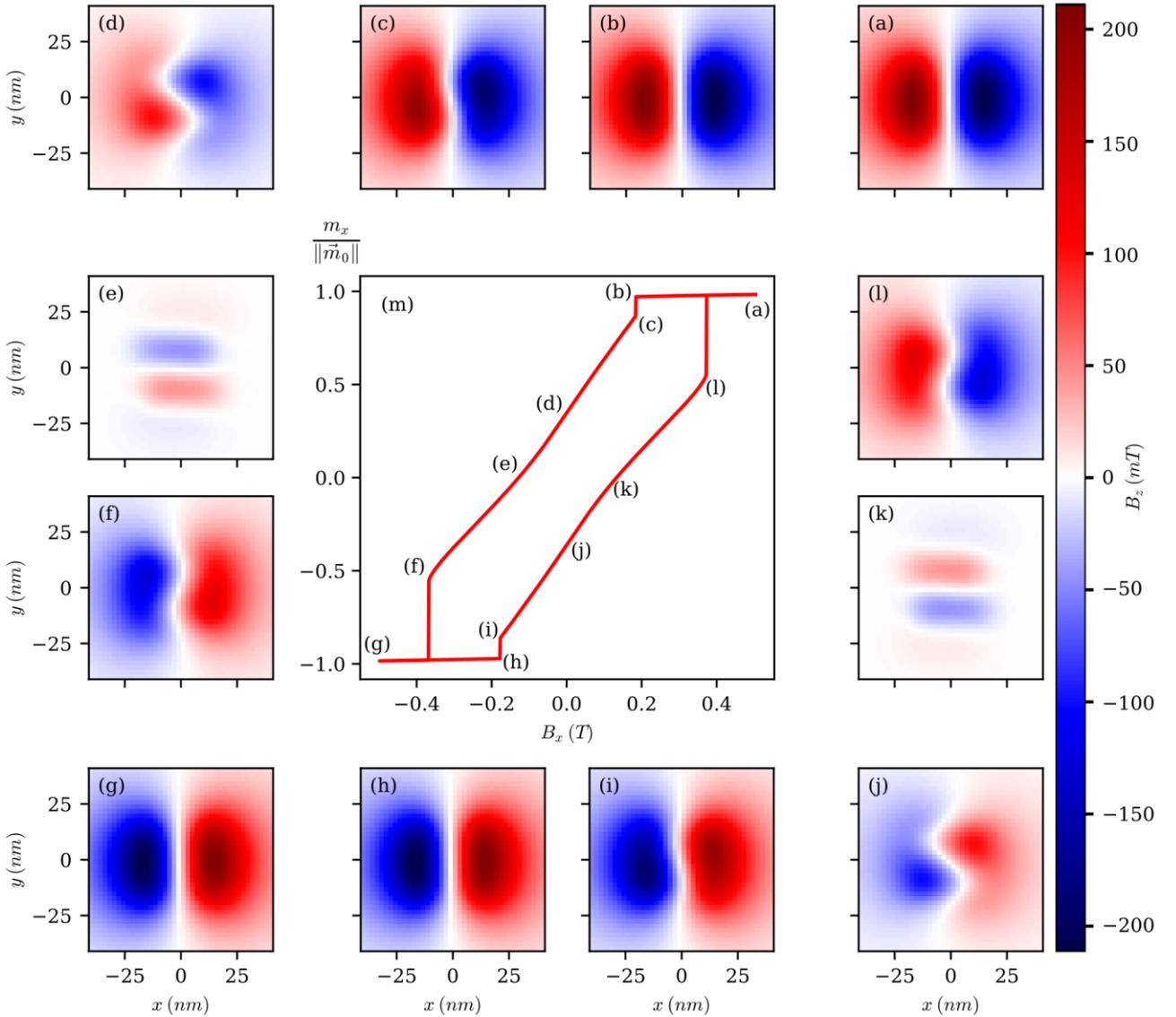

**FIG. S10.** (a-l) Color maps of the normal magnetic stray field of the nanocube. (m) Magnetic hysteresis loop of the 30 $nm$ Fe Nanocube.

# References

[1] S. A. Thiele, J. A. Schaefer, and F. Schwierz, *Modeling of Graphene Metal-Oxide-Semiconductor Field-Effect Transistors with Gapless Large-Area Graphene Channels*, J. Appl. Phys. **107**, 094505 (2010).

[2] J. G. Champlain, *A First Principles Theoretical Examination of Graphene-Based Field Effect Transistors*, J. Appl. Phys. **109**, 084515 (2011).

[3] F. A. Chaves, D. Jiménez, A. A. Sagade, W. Kim, J. Riikonen, H. Lipsanen, and D. Neumaier, *A Physics-Based Model of Gate-Tunable Metal–Graphene Contact Resistance Benchmarked against Experimental Data*, 2D Mater. **2**, 025006 (2015).

[4] L. Petit, T. Fournier, G. Ballon, C. Robert, D. Lagarde, P. Puech, T. Blon, and B. Lassagne, *Performance of Graphene Hall Effect Sensors: Role of Bias Current, Disorder, and Fermi Velocity*, Phys. Rev. Applied **22**, 014071 (2024).

[5] D. C. Elias et al., *Dirac Cones Reshaped by Interaction Effects in Suspended Graphene*, Nature Phys **7**, 701 (2011).

[6] C. Hwang, D. A. Siegel, S.-K. Mo, W. Regan, A. Ismach, Y. Zhang, A. Zettl, and A. Lanzara, *Fermi Velocity Engineering in Graphene by Substrate Modification*, Sci Rep **2**, 590 (2012).

[7] T. Stauber, N. M. R. Peres, and F. Guinea, *Electronic Transport in Graphene: A Semi-Classical Approach Including Midgap States*, Phys. Rev. B **76**, 205423 (2007).

[8] C. W. J. Beenakker and H. van Houten, *Billiard Model of a Ballistic Multiprobe Conductor*, Phys. Rev. Lett. **63**, 1857 (1989).

[9] S. Datta, *Electronic Transport in Mesoscopic Systems* (Cambridge university press, Cambridge, 1995).

[10] A. K. Geim, I. V. Grigorieva, S. V. Dubonos, J. G. S. Lok, J. C. Maan, A. E. Filippov, and F. M. Peeters, *Phase Transitions in Individual Sub-Micrometre Superconductors*, Nature **390**, 259 (1997).

[11] A. K. Geim, S. V. Dubonos, J. G. S. Lok, I. V. Grigorieva, J. C. Maan, L. T. Hansen, and P. E. Lindelof, *Ballistic Hall Micromagnetometry*, Appl. Phys. Lett. **71**, 2379 (1997).

[12] M. Kim et al., *Micromagnetometry of Two-Dimensional Ferromagnets*, Nat Electron **2**, 457 (2019).

[13] L. Theil Kuhn, A. K. Geim, J. G. S. Lok, P. Hedegård, K. Ylänen, J. B. Jensen, E. Johnson, and P. E. Lindelof, *Magnetisation of Isolated Single Crystalline Fe-Nanoparticles Measured by a Ballistic Hall Micro-Magnetometer*, Eur. Phys. J. D **10**, 259 (2000).

[14] F. J. Bonilla, L.-M. Lacroix, and T. Blon, *Magnetic Ground States in Nanocuboids of Cubic Magnetocrystalline Anisotropy*, Journal of Magnetism and Magnetic Materials **428**, 394 (2017).